\documentclass[11pt,preprint]{aastex}

\begin{document}

\def\erg{\hbox{erg}}
\def\cm{\hbox{cm}}
\def\sec{\hbox{s}}
\def\Mpc{\hbox{Mpc}}
\def\keV{\hbox{keV}}
\def\nm{\hbox{nm}}
\def\year{\hbox{yr}}
\def\deg{\hbox{deg}}
\def\arcsec{\hbox{arcsec}}
\def\microJy{\mu\hbox{Jy}}
\def\Hz{\hbox{Hz}}

\def\num{\ifmmode{\nu_m}\else{$\nu_m$}\fi}
\def\nuc{\ifmmode{\nu_c}\else{$\nu_c$}\fi}
\def\nua{\ifmmode{\nu_a}\else{$\nu_a$}\fi}
\def\fnu{\ifmmode{f_\nu}\else{$f_\nu$}\fi}
\def\fnumax{\ifmmode{f_{\nu,m}}\else{$f_{\nu,m}$}\fi}
\def\fradio{\ifmmode{f_{\hbox{\small radio}}}\else{$f_{\hbox{\small radio}}$}\fi}
\def\xie{\ifmmode{\xi_e}\else{$\xi_e$}\fi}
\def\xib{\ifmmode{\xi_b}\else{$\xi_b$}\fi}

\def\tjet{\ifmmode{t_{jet}}\else{$t_{jet}$}\fi}
\def\numax{\num}
\def\nucool{\nuc}
\def\nuabs{\nua}
\def\that{\ifmmode{\hat{t}}\else{$\hat{t}$}\fi}
\def\alphobs{\alpha_{obs}}
\def\kmsMpc{\ifmmode {\rm\,km\,s^{-1}\,Mpc^{-1}}\else
                ${\rm\,km\,s^{-1}\,Mpc^{-1}}$\fi}

\title{Dirty Fireballs and Orphan Afterglows: A Tale of Two Transients}

\author{James E. Rhoads \altaffilmark{1}}

\begin{abstract}
Orphan afterglows are transient events that are produced by
cosmological fireballs and resemble gamma ray burst (GRB) afterglows,
yet are not accompanied by gamma rays.  Such transients may be
produced by jetlike GRBs observed off-axis, and therefore hold great
promise as a test of gamma ray burst collimation.  However, orphans
may also be produced by ``dirty fireballs,'' i.e., cosmological
fireballs whose ejecta carry too many baryons to produce a GRB.

A well designed orphan afterglow search can distinguish between
on-axis dirty fireballs and off-axis orphans in at least two ways.
First, by combining real-time triggers from a wide area, multicolor
search with deeper followup observations, the light curve can be
tracked for a time $\ga 2 t_1$, where $t_1$ is the age of the event at
first observation.  Such a light curve allows simultaneous fits to
$t_1$ and the time decay slope $\alpha$ with sufficient accuracy to
distinguish on- and off-axis orphans. Second, radio followup of
orphan afterglows will show whether the radio flux is falling in time
(as expected for an off-axis orphan) or not (as expected for on-axis
events).  Additional tests involving multi-band monitoring of the
cooling, self-absorption, and $f_\nu$ peak frequencies are also
possible, although much more observationally demanding.

A further complication in orphan searches is that dirty fireballs are
likely to {\it also\/} be collimated, and that collimated dirty
fireballs viewed off-axis will individually be practically
indistinguishable from off-axis GRB afterglows.  To recognize their
presence, orphan afterglow surveys must be sufficiently extensive to
catch at least some dirty fireballs on-axis.
\end{abstract}

\altaffiltext{1}{Space Telescope Science Institute, 3700 San Martin Drive,
 Baltimore, MD 21218} 

\keywords{gamma rays--- bursts}

\section{Introduction}
Orphan afterglows offer great potential as a test of gamma ray burst
collimation (Rhoads 1997).  An orphan afterglow is (conventionally) a
collimated gamma ray burst viewed far enough from the jet axis that no
gamma rays are observed but close enough to the axis that longer
wavelength and less tightly collimated afterglow radiation is
observed.  The properties of such orphans should be similar to the
properties of on-axis afterglows viewed at late times, after the light
curve break expected in collimated GRB afterglows (Rhoads 1999).
Details of the expected orphan afterglow count rate depend
substantially on details of the assumed GRB population.  Pessimistic
numbers result if the kinetic energy in the GRB ejecta is assumed to
scale with jet solid angle (Dalal, Griest, \& Pruet 2002).
Fortunately, present evidence suggests that the burst energy is
approximately independent of collimation angle (Frail et al 2001;
Panaitescu \& Kumar 2002).  In this case, an afterglow will be
detectable to an off-axis angle that is independent of the initial jet
angle, and the orphan afterglow rate can greatly exceed the GRB rate
and provide a strong diagnostic of collimation (Granot et al 2002;
Totani \& Panaitescu 2002; Nakar, Piran, \& Granot 2002 [NPG02]).

However, transients resembling GRB afterglows can also be produced by
cosmological explosions that do not produce gamma ray bursts.
Afterglow emission at any particular frequency reaches peak intensity
at some characteristic time and characteristic Lorentz factor $\Gamma_a$
of the burst ejecta.  The observed behavior is essentially
independent of the initial Lorentz factor $\Gamma_0$ of the explosion
provided that $\Gamma_0 > \Gamma_a$.  Optical depth arguments require
$\Gamma_0 \ga 100$ to produce the observed power law GRB spectra
(Paczynski 1986; Goodman 1986; Krolik \& Pier 1991; Woods \& Loeb
1995.)  Fireballs with $10 \la \Gamma_0 \ll 100$ will not produce
detectable gamma ray emission but will still produce detectable
afterglows at radio, optical, and (if $\Gamma_0$ is large enough)
X-ray wavelengths. 

Thus, dirty fireballs can produce transients with the generic
characteristics of GRB afterglows, i.e., broken power law spectra and
light curves produced by synchrotron emission in a forward external
shock.  Rhoads (2000) remarked on this possibility
and suggested that the difference in light curves could be used to
distinguish the two.  The principle here is that afterglows from
jetlike bursts seen off-axis at frequencies above the peak in $f_\nu$
will show a rapid decay $f_\nu \propto t^{-p}$ (where $p$ is the
electron energy distribution power law index).  In contrast,
afterglows from dirty fireballs will show a much slower decay, $f_\nu
\propto t^{-3(p-1)/4}$ or $t^{1/2-3p/4}$ (depending on whether the
cooling frequency $\nu_c$ lies above or below the observed frequency).

Huang, Dai, \& Lu (2002; hereafter HDL02) examined the requirements for
distinguishing orphan afterglows due to off-axis jets from those due
to dirty fireballs (which they term ``failed gamma-ray bursts'' or
``FGRBS'').  They point out that a major difficulty in distinguishing
the two is the unknown trigger time.  In studies of triggered GRBs,
the decay index $\alpha$ can be determined without ambiguity using the
arrival time of the first gamma rays as the time origin $t_0$.  (A
modest exception can be made for GRB 990123, where the availability of
optical ROTSE data during the GRB itself [Akerlof et al 1999]
introduces some sensitivity of the early decay index to the precise
choice of the time origin.)  However, in an orphan afterglow, the
trigger time is unknown.  All we know is the discovery time $t_d$.
The age of the burst at discovery can be defined as $t_1 \equiv t_d -
t_0$ and becomes parameter of the light curve that we may try to fit.
In general, there is a substantial degeneracy between $\alpha$ and
$t_1$ unless the observations span a period of duration $\Delta t \ga 2
t_1$.  HDL02 suggest a few tests that might help discriminate dirty
fireballs from off-axis GRBs, but conclude that the only promising one
is the possibility (unconfirmed at present) that most dirty fireballs
are accompanied by ``x-ray flashes'' of the type observed by the
BeppoSAX (e.g., Heise et al 2001) and HETE-II satellites.

In this paper, I point out that dirty fireballs observed on-axis may
be identified in orphan afterglow searches through their
multiwavelength properties.  By using quantities whose evolution
changes sign at the jet break, we obtain tests that are robust to the
uncertainty in $t_1$.
Additionally, I examine the distribution of $t_1$ expected under a
simple orphan afterglow search strategy, in order to constrain the
typical monitoring period $\Delta t$ required to reliably fit both
$t_1$ and $\alpha$. 

\section{Multiwavelength Monitoring as a Diagnostic of Dirty
Fireballs} \label{radiosec}
In simple models, an afterglow is characterized by three spectral
break frequencies and a peak flux density (Sari, Piran, \& Narayan 1998).
The breaks are $\num$,
corresponding to the observed peak of synchrotron emission for the
minimum energy of the electron power law distribution; $\nuc$,
corresponding to the observed peak of synchrotron emission for
electrons whose cooling time equals the dynamical age of the
remnant; and $\nua$, the synchrotron self-absorption frequency.
The spectral peak is $\fnumax \equiv \fnu(\num)$.  An
afterglow model predicts the evolution of each of these quantities in
terms of physical quantities that control the fireball evolution and
emission: The initial energy $E_0$, the ambient density $n = \rho /
(\mu m_p)$, and the fractions of the total available energy converted
into relativistic electrons $\xie$ and into magnetic fields $\xib$ at
the forward shock of the expanding blast wave.  Common generalizations
include nonspherical explosions, with initial opening angle $\zeta$,
and a nonuniform ambient medium, with density $n \propto r^{-\delta}$.
(Usually, either $\delta = 0$ for a uniform medium or $\delta = 2$ as
expected for a constant velocity wind from the GRB progenitor).

The light curve of an afterglow depends on the time evolution of the
spectrum in a way that depends on the details of the model (e.g.,
spherical or collimated flow, uniform or wind medium) but in all cases
samples only a fraction of the full information provided by the
multifrequency behavior of the transient.  Thus, when HDL02 examined the
possibilities for distinguishing dirty fireballs from off-axis jets
primarily using the light curve at a single observed frequency, they
used a subset of the available tools and reached unduly pessimistic
conclusions.

Consider a quantity $x$ that behaves as a power law of the time
elapsed since the burst, $x \propto (t-t_0)^{q}$.  Following HDL02,
suppose we do not know $t_0$ but have made a guess $t_0^{obs} = t_0 +
T$ that is in error by amount $T$.  Then $x \propto (t - (t_0^{obs} -
T))^{q}$.  Without loss of generality, we can set the origin of time
to the assumed burst trigger, $t_0^{obs} = 0$, simplifying the
previous expression to $x\propto (t+T)^{q}$.  Then the instantaneous
slope measurement becomes $q^{obs}(t) \equiv d\log{x}/d\log(t) = q t /
(t+T)$.  While our measurement of $q$ will be substantially in error
unless $|T|/t \ll 1$, we see that the sign of $q^{obs}$ is
unaffected.  (Note that $(t+T)$ is always positive, since our first
observation cannot precede the actual trigger time.)

We should therefore measure the behavior of one or more quantities
whose sign changes at the transition from effectively spherical to
effectively jetlike behavior.  Table~1 lists several potentially
applicable quantities, together with their expected evolution after
the jet break and before the break in either a uniform density ambient
medium or a wind environment.  Spherically symmetric GRB remnants
expanding into a wind density profile $\rho \propto r^{-2}$ are more
similar to jetlike afterglows than are their uniform density
counterparts in terms of several observables (peak flux density,
absorption frequency, and fixed frequency light curves).

\begin{deluxetable}{lccc}
\tablecolumns{4}
\tablewidth{250pt}
\tablecaption{Indicators of Orphan Afterglow Origins}
\tablehead{
\colhead{Quantity} & \multicolumn{3}{c}{Conditions:}\\
 & \colhead{{$t<\tjet$}} & \colhead{$t<\tjet$}
 & \colhead{$t>\tjet$}\\
 & \colhead{uniform} & \colhead{wind} & }
\startdata
\fnumax & 0 & -1/2 & -1 \\
\nucool & -1/2 & +1/2 & 0 \\
\nuabs & 0 & -3/5 & -1/5 \\
\fradio\tablenotemark{a} & 1/2 & 0 & -1/3 \\
\enddata
\tablenotetext{a}{``\fradio'' is here defined as the flux density for
frequencies $\nu$ such that $\nuabs < \nu < \numax$.  This corresponds
to cm-wave and mm-wave radio frequencies for typical GRB afterglows.}
\tablecomments{\small
Power law exponents for the time evolution of afterglow
emission parameters.  Each line of the table lists a physical quantity
``$x$'' (in the first column) followed by its time evolution exponents
$q$ (defined as $x \propto t^q$) in three different evolution regimes.
Columns 2 and 3 are both for an on-axis afterglow viewed before the
jet break time, which corresponds to the early time regime for
conventional GRB afterglows and dirty fireballs.  Column~2 is for a
uniform ambient density and column~3 for $\rho \propto r^{-2}$ (as
expected if the ambient medium is a steady wind from the
GRB progenitor).  Column 4 is for the evolution after the jet break
(independent of ambient density profile; see Chevalier \& Li 1999),
and so illustrates the full observable evolution of an orphan
afterglow from an off-axis jet.  These physical parameters were chosen
because they change sign at $t\approx \tjet$, at least for some
ambient density laws.  They therefore provide robust tests of whether
an observed orphan afterglow is due to an off-axis event or an on-axis
``dirty fireball.''  The tabulated time evolution slopes are taken
from Rhoads (1999), Chevlier \& Li (1999, 2000), and Sari, Piran, \&
Halpern (1999).}
\end{deluxetable}

The most promising observable is the radio light curve, for two
reasons.  First, the radio flux falls with time after a jet
break, while it remains flat or rises prior to a jet break
for any likely density profile.  Second, it is the simplest to observe,
requiring data at only one frequency.

Combinations of the remaining diagnostics can be used similarly to
distinguish among the three broad model classes in table~1.  In
particular, the cooling frequency $\nucool$ can be combined with the
spectral peak $\fnumax$.  If $\fnumax$ is declining with time, it
indicates either a wind-like ambient medium or a post-break jet.  In
the first case, $\nucool$ should rise with time, while in the second
case $\nucool$ should remain steady.  The self-absorption frequency $\nuabs$
could substitute for $\fnumax$ in this argument, though its time
dependence is weaker than that of $\fnumax$.  These two-diagnostic
tests will be observationally challenging, because the observations
involved require good measurements over a wide wavelength range and
because it may be hard to distinguish between a truly non-evolving
$\nucool$ and a relatively weak evolution in an old orphan afterglow.
Thus, we tend to prefer either the radio light curve or a suitably
detailed optical/IR light curve (section~\ref{fits}) for their
conceptual and observational simplicity.

\section{Two-parameter Fitting of Optical Orphan Light Curves}
\label{fits}
Even in the case where multiwavelength data are not available, the
nature of an optically discovered orphan may be determined by
sufficiently accurate monitoring observations.  There are three
relevant adjustable parameters of the monitoring strategy: Time
sampling $\delta t$, number of epochs observed $N$,
and photometric accuracy.  We will presume in this section a
monitoring strategy where a wide-field survey camera of moderate
sensitivity is used to find orphan afterglow candidates, with each
epoch of data processed in a time $\ll \delta t$.  The search strategy
should employ multiple optical filters spanning the Balmer break, so 
that color information can be used to identify the best 
orphan afterglow candidates in a single epoch (Rhoads 2001).  Candidates thus
found could then be followed up by larger telescopes.  We will
therefore presume that the fractional uncertainty in flux measurements
is constant throughout the monitoring campaign.  We will write this
accuracy in magnitudes, $\delta m = \delta (2.5 \log_{10}f) \approx
1.0857 (\delta f) / f$, where $f$ and $\delta f$ are the flux and its
associated uncertainty in linear units.

We now ask what kind of observational strategies will be sufficient to
distinguish the light curves of an off-axis orphan afterglow and an
on-axis dirty fireball.  Assume that we either obtained the spectral
slope $\beta$ of the transient already (since our initial search
exploits color information) or else will obtain it during the followup
campaign.  Further, assume that we can deredden the measured spectral
slope moderately well using its observed spectral curvature.  (This is
not trivial, but the required accuracy is not high.)  Then we can
infer the expected light curve slopes under either of two possible
regimes.  If the observations are above the cooling frequency for the
burst, the electron spectral slope index $p$ is given by $p = 2 \beta$
and the time decay slope is either $p=2\beta$ (for an off-axis orphan)
or $3p/4 - 1/2 = 3\beta/2 - 1/2$ (for an on-axis dirty fireball).  On
the other hand, if the observations are below the cooling frequency,
we have $p = 2 \beta + 1$ and decay slopes of either $p = 2\beta+1$ or
$3p/4 - 3/4 = 3 \beta/2$.  To distinguish the two afterglow regimes,
we therefore need to tell apart two time decay slopes $\alpha$
differing by either $p/4+1/2 = \beta/2 +1/2$ or by $p/4 + 3/4 =
\beta/2 + 1$.  Since $p \approx 2$ is typical we will be satisfied
with any observational strategy that can measure $\alpha$ to a
one-sigma accuracy of $\pm 0.3$, sufficient to distinguish the above
two scenarios at a $3\sigma$ level.

To determine the error in $\alpha$, I have generated artificial power
law light curves sampled according to an assumed observing strategy
and with artificial noise added.  I then fit each simulated light
curve with a model $f = f_1 [( \that + t_1) / t_1]^{-\alpha}$,
where \that\ is the
time elapsed since the first observation.  The first numerical
experiments were used to demonstrate that the accuracy of the
recovered spectral slope $\alphobs$ does not vary substantially with
the intrinsic slope $\alpha$ over the range $1 \la \alpha \la 3$ for a
wide range of observing strategies.  The remaining simulations therefore
used a uniform value of $\alpha = 1.8$, near the midpoint of the
interesting range.

Because we model the light curves to be power law decays after the
first detection at time $t_1$, there is only one characteristic
timescale in the light curve ($t_1$).  Only the ratio $\delta t / t_1$
need be considered in studying the accuracy of light curve fitting.  I
have therefore fixed $\delta t = 1$ day in the simulations while
allowing $t_1$ to vary.  Results for $\delta t = k$ days can be
inferred by replacing $t_1$ with $t_1 / k$ in the figures.

\begin{figure}[ht]
\plotone{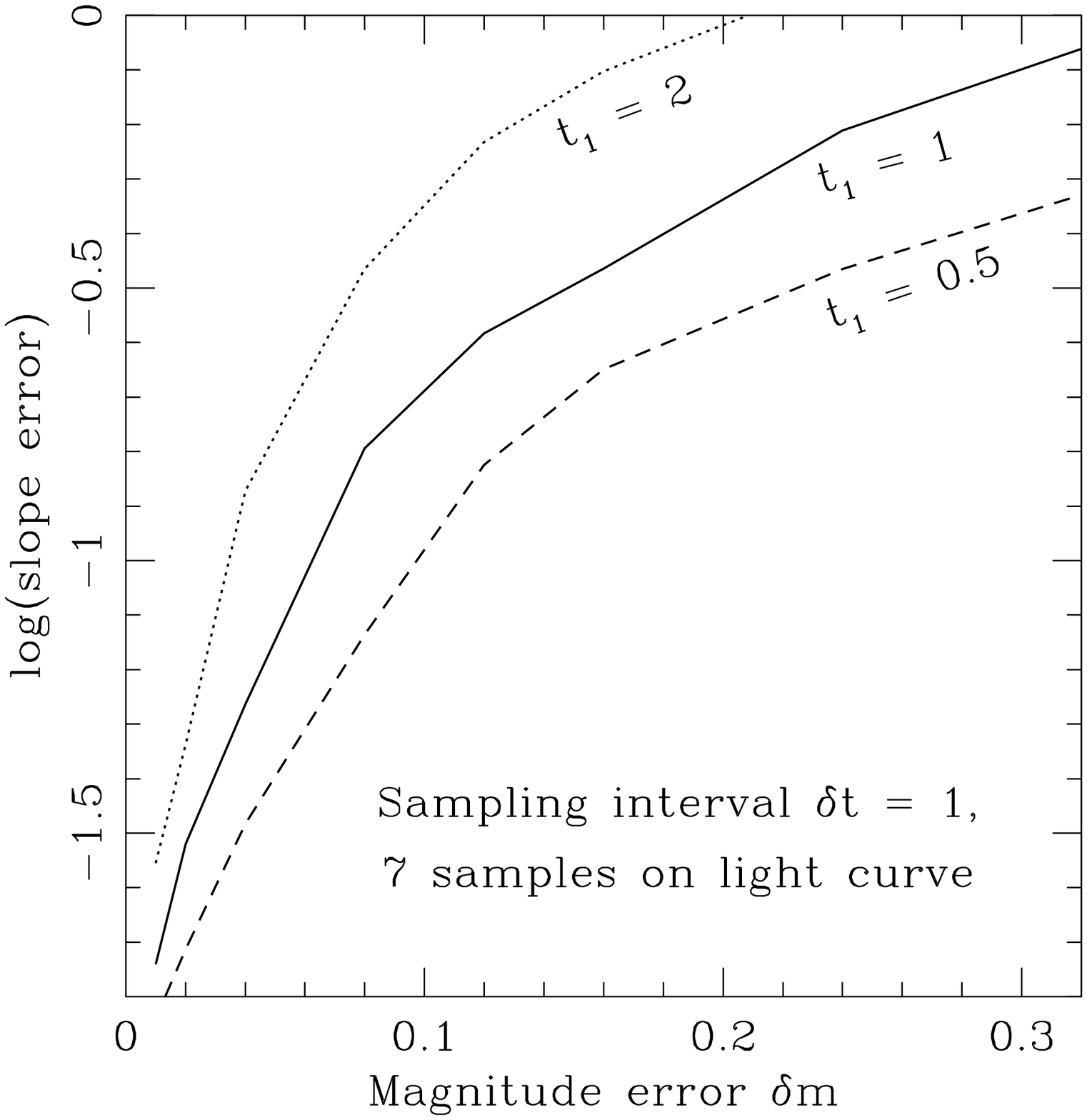}
\caption{The error in the measured light curve decay slope for an
orphan afterglow.  The decay is taken to be a pure power law with an
unknown time origin that is fitted as a parameter of the model.
Curves are shown for three different ages of the event at the time of
first observation. Observations are assumed at 1 day intervals
continuing for 7 observations in total.  All observations are assumed
to have the same logarithmic flux error, which is plotted in magnitude
units on the $x$-axis.  The resulting error in inferred slope, shown
on the $y$-axis, is derived by Monte Carlo simulations of a least
squares fitting procedure to the light curve.  A logarithmic slope
error of $-0.5$ is approximately sufficient to distinguish orphan
afterglows produced by on-axis dirty fireballs from those produced by
off-axis gamma ray bursts. \label{merr}}
\end{figure}

Figure~\ref{merr} shows how the uncertainty in $\alpha$ varies with
photometric accuracy, for a baseline strategy of nightly samples ($\delta
t = 1$ day) continued for 7 nights total and for $t_1 = 0.5, 1, 2$ days.

\begin{figure}[ht]
\plotone{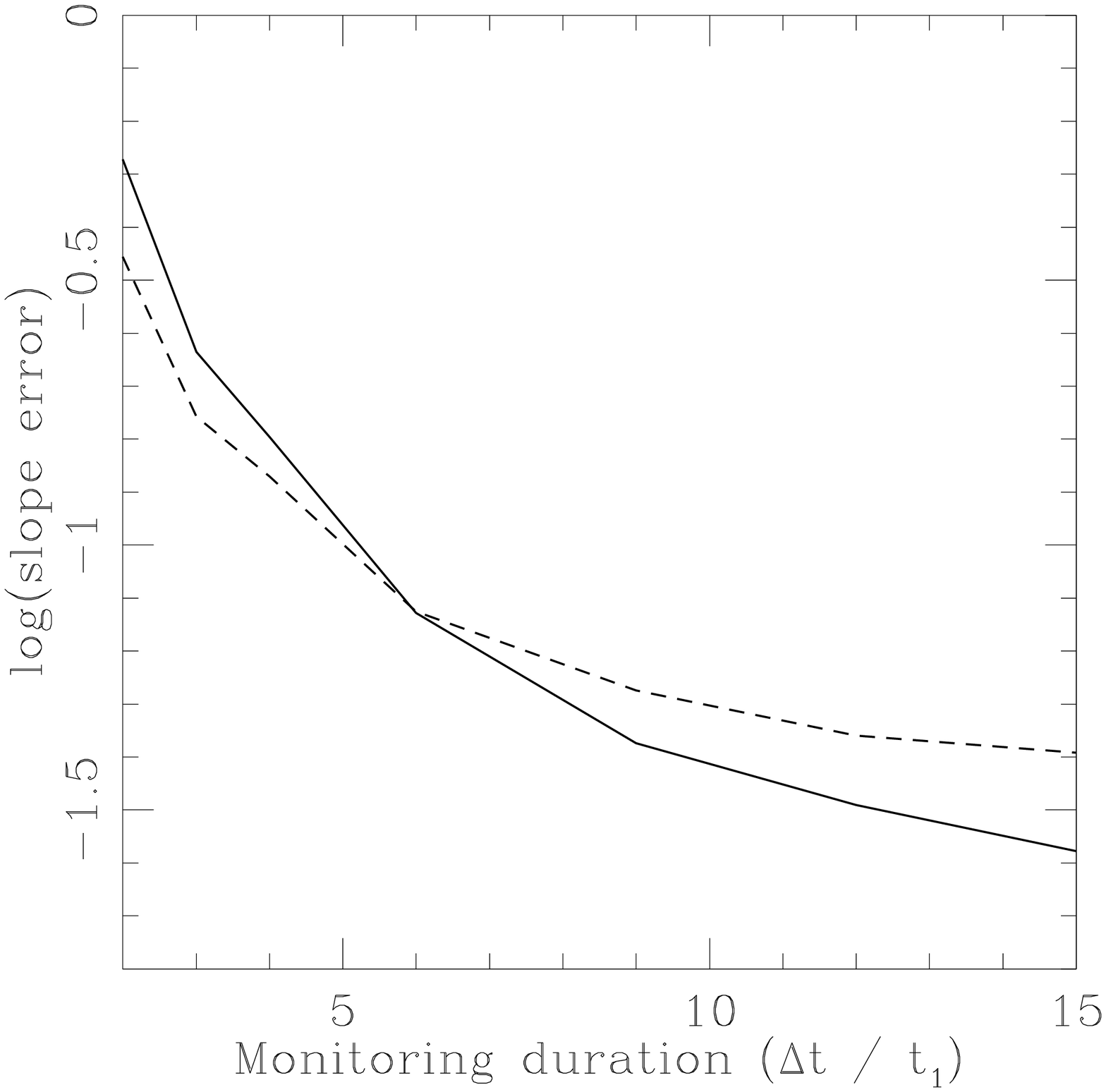}
\caption{The error in the light curve decay slope as a function of the
monitoring campaign duration.  The solid line shows the result if the
number of samples is varied and the interval remains fixed (with
$\delta t = t_1$).  The dotted curve shows the result for a fixed
number of data points ($N=7$) and a varying $\delta t$.  The flux
error on each measurement is taken to be $\delta m = 0.05$ for both
curves.
\label{dur}}
\end{figure}

Figure~\ref{dur} shows the variation of $\delta \alpha$ with the total
duration $\Delta t = (N-1) \delta t$ of monitoring, for the two cases
where we (a) fix $\delta t / t_1 = 1$ and vary $N$, and (b) fix
$N=7$ and vary $\delta t / t_1$.  The observational error on each
measurement is held constant at $\delta m = 0.05$ throughout.

The uncertainties in the measured values of $\alpha$ and $t_1$ are
strongly correlated.  The slope of the correlation can be estimated
easily by noting that
\begin{equation}
{d \log f \over d t} = {-\alpha \over {\that + t_1}} ~~.
\end{equation}
This implies that the expected error $\delta t_1 \approx -\delta \alpha /
(d\log f/dt)$, where we can evaluate the observable quantity $d\log f
/ dt$ near the midpoint of the observations.

\section{The Distribution of Discovery Times}
The age of an orphan afterglow at first discovery is a key factor in
how well we can determine its decay slope.  In this section we predict
this distribution in a simple orphan afterglow model.  Our derivation
is closely analogous to the Nakar et al (2002) orphan rate
calculation, and includes the approximation that GRB jets have a 
standard energy.  In this model, the light curve of off axis
orphan afterglows assumes a standard form.  The maximum time $t_{max}(z,m)$ at
which an orphan is detectable is then determined uniquely by the
redshift $z$ of the burst and the magnitude limit $m$ of
the observations.  For completeness, we reproduce here the result
of inverting equations 6--7 of NPG02:
\begin{equation}
{t_{1,max} \over \hbox{day}} = \left\{460{g_0(p) \over g_0(2.2)} 10^{2.2-p \over 4}
  \epsilon_{e,-1}^{p-1} \epsilon_{B,-2}^{(p-2)/4} n_0^{-(p+2)/12}
  E_{0,50.7}^{(p+2)/3} \nu_{14.7}^{-p/2} (1+z)^{p+2 \over 2}
  d_{L,28}^{-2} {\microJy \over f_{min}} \right\}^{1/p}
\end{equation}
for the case where $\nu_c < \nu$, and
\begin{equation}
{t_{1,max} \over \hbox{day}} = \left\{170{g_1(p) \over g_1(2.2)} 10^{2.2-p \over 4}
  \epsilon_{e,-1}^{p-1} \epsilon_{B,-2}^{(p+1)/4} n_0^{(3-p)/12}
  E_{0,50.7}^{(p+3)/3} \nu_{14.7}^{(1-p)/2} (1+z)^{p+3 \over 2}
  d_{L,28}^{-2} {\microJy \over f_{min}} \right\}^{1/p}
\end{equation}
for the case $\nu_c > \nu$.  Quantities in these equations are scaled
to dimensionless or cgs unit values whose logarithms are indicated in
the subscripts: $p$ is the slope of the power law distribution of
energies for electrons accelerated in the forward shock of the
expanding GRB remnant; $\epsilon_{e,-1}$ and $\epsilon_{B,-2}$ are the
fractions of internal energy going into relativistic electrons and
magnetic fields at this shock, scaled to units of $0.1$ and $0.01$
respectively; $n_0$ is the number density of the ambient medium;
$E_{0,50.7}$ is the initial kinetic energy of the ejecta in units of
$10^{50.7} \erg$, $\nu_{14.7}$ is the observed frequency in units of
$10^{14.7}\Hz$, $z$ is the GRB redshift, and $d_{L,28}$ is the
luminosity distance in units of $10^{28} \cm$.  The functions
$g_0(p) =10^{-0.56 p} (p-0.98) [(p-2)/(p-1)]^{p-1}$
and $g_1(p) = 10^{-0.31 p} (p-0.04) [(p-2)/(p-1)]^{p-1}$
vary slowly with $p$ for $p \ga 2.2$ but
become undefined for $p \le 2$, since the assumed physical model for
the radiating electron population gives a divergent total electron
energy if $p \le 2$.  

\begin{figure}[ht]
\plotone{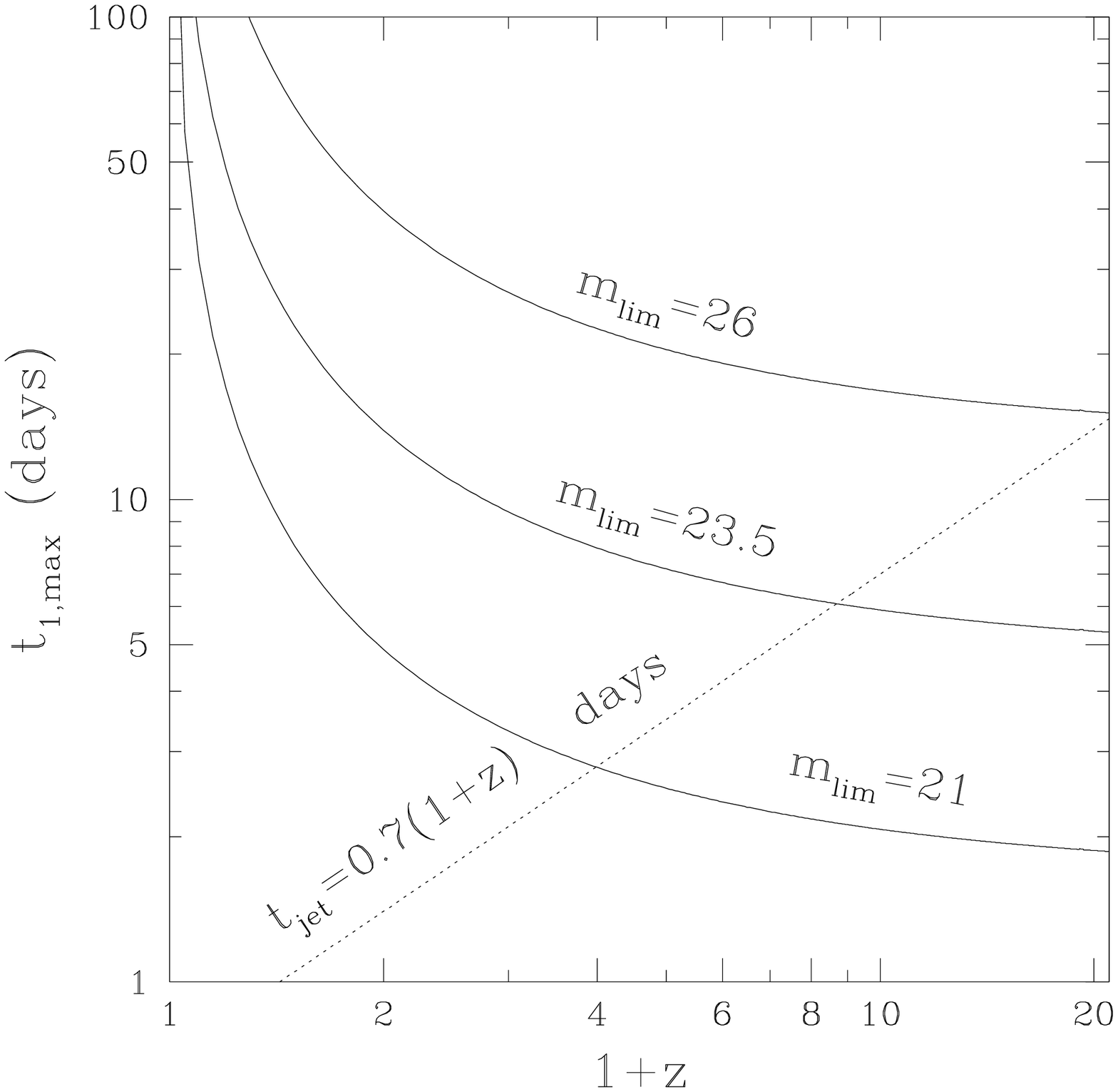}
\caption{The maximum age at which an afterglow could be detected
as a function of redshift, for three different survey flux thresholds.
The light curve assumed is the ``universal'' late time light curve of
an off axis jet, and the plotted curves are therefore not meaningful
for values of $t_{1,{\rm max}} < \tjet$.  The fluxes are derived
from equation~2, with all physical parameters of the fireball held at 
their fiducial values, so that $t_{1,max} = \left[ (1+z)^{p+2\over 2}
d_{L,28}^{-2} \left(460 \microJy \over f_{min}\right) \right]^{1/p}$ with
$p=2.2$.
\label{tmax}}
\end{figure}

An interesting feature of $t_{1,max}$ is that it declines extremely
slowly at high redshifts.  $t_{1,max}(z,m)$ is plotted for a few
values of $m$ in figure~\ref{tmax}.  Because our calculation of 
$t_{1,max}$ has assumed the ``universal'' light curve for an off-axis
jet, it is only valid for $t_{1,max} > \tjet$.  We therefore overplot
$t_{jet,obs}$ for rest-frame $\tjet = 0.7$ day.

Consider first the distribution of ``turn-on time''\footnote{The time
at which an orphan afterglow could first be detected assuming continuous
monitoring.}  $t_{on}$ under the approximation that the afterglow
becomes visible when $\theta_{obs} = 1 / \Gamma$.  In the late time
evolution of a spreading GRB jet, $\Gamma \propto t^{-1/2}$ (Rhoads
1997, 1999).  The distribution of off-axis angles should go as
$dN_{GRB} / d\theta_{obs} \propto \theta_{obs}$.  If we consider a
single redshift and a single jet energy, we then obtain
\begin{equation}
\left(dN_{GRB} \over dt_{on}\right)_{z,E_0} \propto 
\left\{
\begin{array}{ll}
  1 / (1+z) & \hbox{  if }\tjet < t_{on} < t_{1,max}(z,E_0) \\
  0 & \hbox{otherwise}
\end{array}
\right. ~~.
\end{equation}
Here $N_{GRB}$ has units of events per day per unit volume, while
$dN_{GRB} \over dt_{on}$ is in events per day$^2$ per unit volume.
The factor of $(1+z)^{-1}$ in the rate is required to account for 
cosmological time dilation of the observed turn-on time.

If we now assume a search strategy with observations regularly spaced
at interval $\delta t$, the above distribution is modified to yield the
distribution of age $t_1$ at detection:
\begin{equation}
\left(dN_{GRB} \over dt_{1}\right)_{z,E_0} \propto 
\left\{
\begin{array}{ll} \min(t_{1} / \delta t, 1) / (1+z)
 &  \hbox{  if } \tjet < t_{1} < t_{1,max} \\
0 & \hbox{otherwise}
\end{array} \right. ~~.
\end{equation}
In the limit of continuous observations, $\delta t \rightarrow 0$, the
two expressions become identical as required.

Now let us integrate over redshift.  The rate of orphan detections
per unit $t_1$ becomes
\begin{equation}
R_{orph}(t_1) = \int_0^{z_{max}} \left(dN_{GRB} \over dt_{1}\right)_{z,E_0}
{dV/dz \over 1+z} dz
\end{equation}
where $m$ is the magnitude limit of the survey.  Again following Nakar
et al (2002), we take $n(z)$ to grow as $n(z) \propto 10^{0.75z}$ for
$z\le1$ and hold $n(z)$ constant for $z \ge 1$.  We fix the cosmology
to currently fashionable values, i.e., $H_0 = 70 \kmsMpc$, $\Lambda_0
= 0.7$, $\Omega_0 = 0.3$.  Numerical integration of $R_{orph}(t_1)$
was done using the ``Libcosm'' C library (Yoshikawa 1999).

The overall GRB rate density was normalized to set the total on-axis
GRB rate over the redshift range $0<z<5$ to about 900 GRB/year, using
a fixed rest-frame $\tjet = 0.7$ day. This matches reasonably well the BATSE
rate. A more accurate normalization may be impossible given present
observational uncertainties.  The shape of the $t_1$ probability
distribution is independent of the exact normalization of the GRB
event rate.

\begin{figure}[ht]
\plotone{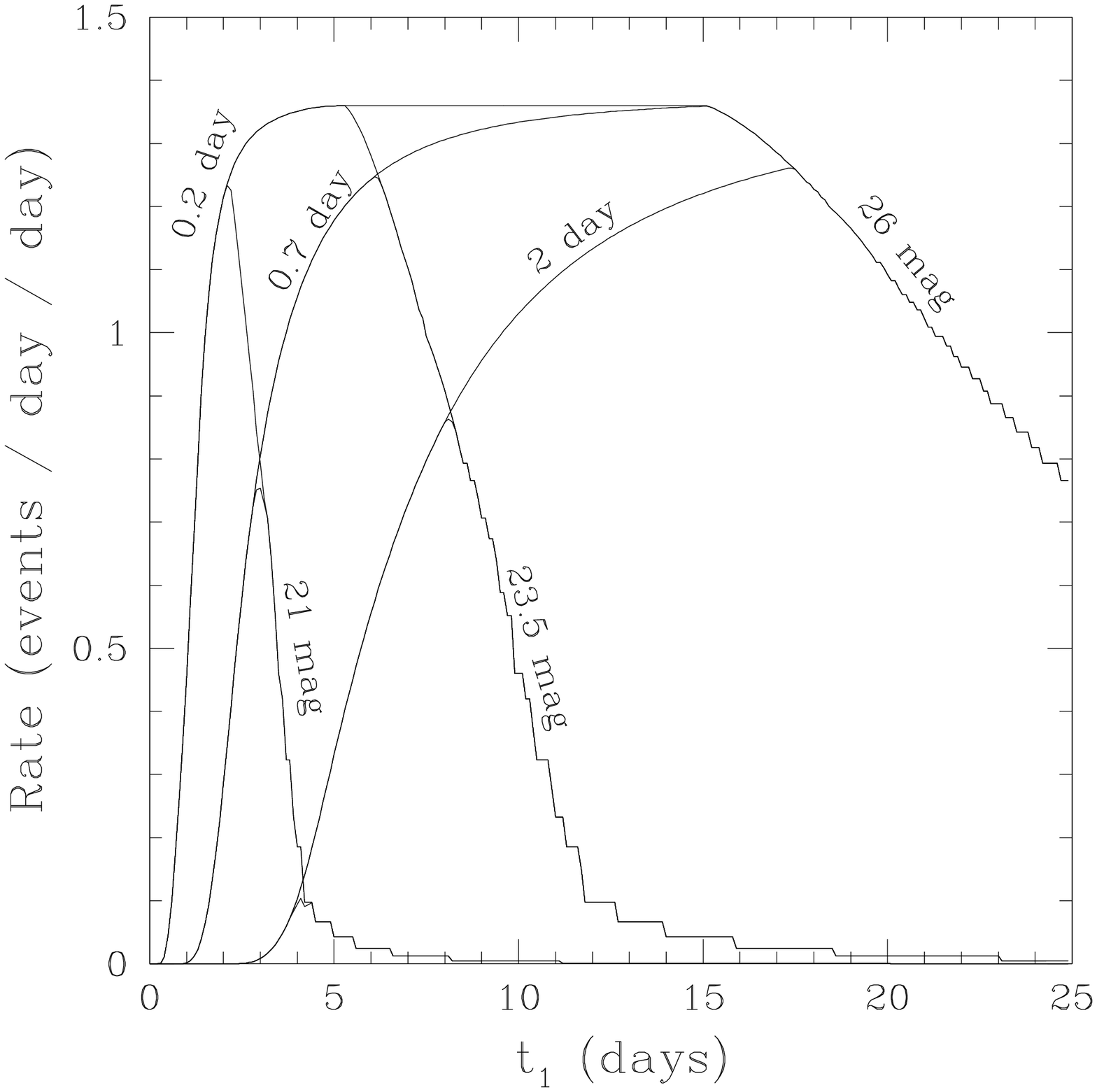}
\caption{Distributions of orphan afterglow age at the time of first
detection for spreading jet models with a range of jet break times and
monitoring programs.  The rising part of each curve is
effectively set by the jet break time (marked on each curve) and the
light curve sampling interval (here fixed at 1 day).  The declining
part of each curve is effectively set by the magnitude limit of the
survey.
\label{t1dist}}
\end{figure}

Results of our numerical $R_{orph}(t_1)$ calculations are shown in
figure~\ref{t1dist}.  A typical rate curve will show either three or
four segments.  The first is a rising segment, determined by the jet
break time and the observational time sampling interval.  The second
is a ``plateau'' where $R_{orph}(t_1)$ is essentially constant.  This
occurs when $t_1$ exceeds both the sampling interval $\delta t$ and
the jet break time $\tjet$, and simultaneously the sensitivity of the
survey is sufficient to detect orphans up to very high redshift.  Such
conditions are only met for comparatively sensitive surveys and short
jet break times.  The next segment is a falling one, whose typical time
scale is determined by the magnitude limit of the survey.  The last
component is a low-level, late time tail comprised of the closest
bursts, which can be detected at large angles off axis but over a very
small volume.

We see that the peak of $R_{orph}(t_1)$ occurs on times ranging from a
couple of days to a couple of weeks, depending on the characteristic
jet break time and the survey magnitude limit.  Moreover, it can be
quite sharply peaked for surveys of relatively bright flux limits
($\sim$ 21st magnitude) but is rather broad for more sensitive
surveys.

\section{Implications for Orphan Afterglow Survey Strategy}
I now outline an observing strategy for optical regime orphan
afterglow surveys that incorporates the results of preceding sections
together with earlier studies.

Monitoring should be carried out to a depth sufficient to find
afterglows that are at least a few days old when their first photons
reach the observer.  Based on figure~\ref{t1dist}, this implies a
depth of at least 23rd to 24th magnitude.  Monitoring should also be
carried out in multiple filters spanning at least the range $400 \nm <
\lambda < 800 \nm$, in order to successfully separate afterglows from
other transient events in a single epoch (Rhoads 2001).  This will
then allow followup observations to considerably greater depths, which
may be needed as the event fades below the detection threshold of the
original search.  Tbe search cadence for monitoring should be
reasonably short, $\delta t \la \tjet \sim 1 \hbox{day}$, in order to
distinguish jetlike events seen on-axis from other afterglows (see
section~\ref{discuss}).

Once a candidate has been identified, optical followup observations
should proceed with a cadence scaled to the expected age of the event
at first detection.  For a search to 23rd magnitude, every 2 nights is
probably sufficient, while shallower searches should be followed every
night initially.  Deeper searches might require only one or two
samples per week.  Such monitoring, pursued for half a dozen epochs
(corresponding to a monitoring time $\Delta t \ga 2 t_1$) would allow
a sufficiently accurate measurement of the decay slope $\alpha$ to
distinguish between on- and off-axis events, despite the uncertainty
in trigger time.

Each identified orphan afterglow candidate should also be observed
twice or more at radio wavelengths.  This will allow the test of
section~\ref{radiosec} to be applied and so distinguish between a
laterally spreading jet and an on-axis dirty fireball.

Supplemental tests based on the nature of the transient's host galaxy
should also be applied, given the history of orphan afterglow searches
so far.  Some active galactic nuclei are capable of producing flares
that closely resemble orphan afterglows, and indeed the first
published optical orphan afterglow candidate (Vanden Berk et al 2002)
turned out to be just such an event (Gal-Yam et al 2001).  Such AGN
flares can be separated from cosmological fireballs by long-term
monitoring and/or spectroscopy of the host galaxy.

While the above strategy is observationally demanding, it should be
within reach of already planned instruments.  The ARAGO project (Bo\"{e}r
2001) will operate a transient search using a 1.5m telescope
with a wide field ($2\times 2^\circ$) camera.  Such a system will reach an
interesting depth for off-axis orphan searches and may provide a few
detections per year for followup with larger instruments.  Further
away but more powerful is the Large Synoptic Survey Telescope
(LSST)\footnote{http://www.lssto.org},
which will survey about half the sky to a depth of $\sim 24$ mag and a
cadence of $\sim 1$ week.  Such a survey could discover hundreds of
orphan afterglows per year.  It would also provide real-time triggers,
since multiple filters are planned.  After a year or two of operation
it would provide a good monitoring baseline for all afterglow
candidates in its survey region, allowing comparatively easy rejection
of some AGN.  Finally, since the LSST will be an 8m class telescope,
it would be capable of performing its own very deep followup
observations, provided that enough flexibility in scheduling exists to
spend extra integration time on interesting transients identified in
the course of the survey.

\section{Discussion} \label{discuss}
It is likely that dirty fireballs, if they exist, are collimated by
whatever mechanism collimates the gamma ray bursts.  This raises the
likelihood (discussed also by HDL02 and NPG02) that we will observe orphan
afterglows due to off-axis dirty fireballs.
It will be very difficult to determine whether any particular off-axis event
emitted gamma rays.  The best possibility here is to look for
reprocessed radiation in some form, e.g., iron emission lines.  Such
reprocessed radiation may be substantially more isotropic than the
synchrotron radiation from relativistic ejecta, and indeed has been proposed
as a calorimeter of GRB energetics (Ghisellini et al 2002).
Ghisellini et al point out that X-ray emission lines may still be
anisotropic due to optical depth effects.  Fortunately, the
observation of an orphan afterglow implies that we are not terribly
far from the jet axis.  Observationally, searches for X-ray lines
would require very rapid and sensitive X-ray followup of orphan afterglows
discovered in ongoing variability surveys, given that the observed
lifetime of all X-ray lines so far observed is $\la 2$ days and the
typical time lag to observe an orphan is comparable.
Additionally, X-ray lines only require photons sufficiently
energetic to excite the iron K edge, $\sim 7 \keV$.  Thus, while these
lines would place an interesting constraint on the fireball Lorentz
factor, they do not guarantee that an event produced gamma rays.

Predictions for orphan afterglows will also depend on the GRB jet
profile, that is, the variation of ejecta Lorentz factor
($\Gamma(\theta)$) and kinetic energy ($dE/d\Omega(\theta)$) with the
off-axis angle $\theta$.  Our calculations in this paper have followed
the expanding top-hat approximation, where both $\Gamma$ and
$dE/d\Omega$ are step functions with some fixed value at $\theta <
\theta_j$, dropping to zero at $\theta > \theta_j$.  This is the best
explored model for collimated GRB afterglows, including essentially
all work on orphan afterglows to date.  However, it remains an
idealized case of more general and realistic jet models.  The light
curve breaks expected in an expanding top-hat case (Rhoads 1999) are
expected also in cases where $dE/d\Omega \propto \theta^{-k}$ with $k
\approx 2$.  However, in this case the break time indicates the
observer's off-axis angle $\theta_{obs}$ instead of a jet opening
angle $\theta_j$ (Rossi et al 2002; Zhang \& M\'{e}sz\'{a}ros 2002).
Similarly, while orphan afterglows are a reasonable expectation in
such ``inhomogeneous'' jets, the detailed predictions may change.  If
we allow $dE/d\Omega$ to decline smoothly with $\theta$ while holding
$\Gamma$ fixed, we expect most afterglows to be accompanied by
$\gamma$-ray emission, since there are fast ejecta along the line of
sight regardless of $\theta_{obs}$.  On the other hand, if we fix
$dE/d\Omega$ and allow $\Gamma$ to decline, we enter a regime where
events viewed near the jet axis yield GRBs and those seen at larger
angles behave as on-axis dirty fireballs.  The fully general case
(where both $dE/d\Omega$ and $\Gamma$ vary smoothly) may yield
a range of behaviors between these limiting extremes.

A final complication is the need to correctly account for the
population of conventional GRB afterglows in orphan afterglow
searches.  Limits to the sky coverage and sensitivity of GRB detectors
imply that the $\gamma$-ray emission will go undetected in a
substantial fraction of the conventional GRB afterglows discovered by
any orphan search.  Being on-axis, these events will generally show a
slow initial decay followed by a jet break.  If the time sampling of
the orphan survey is sufficiently dense ($\delta t < \tjet$), these
conventional afterglows will typically also be seen with brighter
initial magnitude than off-axis orphans.  Thus, they resemble the
likely properties of a dirty fireball population, and the
observational tests we have described will serve to distinguish them
from off-axis orphans.

Confusion between conventional GRBs and dirty fireball populations
remains possible.  To remove this confusion, the expected rate of
conventional afterglows in an orphan survey could be calculated using
the known $\gamma$-ray and afterglow properties of conventional GRBs.
An accurate correction might be difficult at present, because the
ratio of afterglow to $\gamma$-ray flux shows a wide dispersion and
because the data in currently available samples are very heterogeneous.
Fortunately, more homogeneous and statistically tractable samples
should become available with the launch of the Swift mission.

Swift offers another potential way to mitigate confusion by
conventional GRBs: Orphan searches could be conducted preferentially
in regions monitored by the Swift Burst Alert Telescope (BAT).  With a
set of $\sim 10$--$20$ fields evenly spaced on the sky, at least one
would always fall within the 2 steradian field of the BAT.  The BAT
duty cycle at any point on the sky is limited by Earth occultation to
$\la 50\%$, with each continuous observation $\la 50$ minutes.  For
each orphan search field, the number of afterglows with Swift
counterparts could be counted, and the total number of conventional
afterglows in the sample could then be inferred by a simple correction
for the Swift duty cycle.  This correction could be fairly precisely
derived using the known Swift pointing history, and would a factor
between $2$ and $10$.  

On-axis orphan searches could be performed with a much wider field,
brighter flux limit, and faster observing cadence than the off-axis
searches we have primarily considered here (Nakar \& Piran 2002).
Such surveys are natural successors to current robotic telephoto lens
projects (e.g., Kehoe et al 2002).  In this limit, it would be
possible to {\it always\/} monitor fields in the BAT field of view,
and thereby know whether any particular optical flash was or was not
accompanied by a GRB.  Implementing this strategy would require
switching fields approximately twice per Swift orbit (assuming Swift
changes targets to avoid Earth occultation), and would require
real-time knowledge of the Swift pointing, but neither of these should
be an insurmountable difficulty.

A general and robust orphan search strategy is to require sufficient
areal and time coverage to ensure the detection of some conventional
GRB afterglows, and hence also of on-axis dirty fireballs from any
population with an event rate comparable to the GRB rate.  Tbe search
cadence should be kept at $\delta t \la \tjet$ (suggesting nightly
observations) in order to catch the on-axis orphans before their jet
break and so recognize them.  If such coverage fails to find on axis
orphans, it will mean that dirty fireballs do not contribute
substantially to the overall orphan afterglow rate.  More probably, we
will find on-axis orphans.  Comparing their rate to the GRB rate will
tell us the fraction of cosmological fireballs that achieves $\Gamma_0
\ga 100$.  A large class of dirty fireballs might be found, and if it
is, would tell us that GRBs are a minority population in a much larger
class of cosmological fireballs.  The combined knowledge of the
on-axis dirty fireball rate and the off-axis orphan rate would then
allow correction of the orphan statistics for dirty fireballs, and
lead to the desired constraint on GRB collimation from orphan counts.

\acknowledgements 
I thank an anonymous referee for useful comments that strengthened
this paper.  This research was partially supported by an Institute
Fellowship at The Space Telescope Science Institute (STScI).

\end{document}